\documentclass[a4paper,reprint,5p,times,twocolumn,10pt,twoside]{elsarticle}
\usepackage{amsmath,amsthm,amscd,color}
\usepackage{amsfonts}
\usepackage{amssymb}
\usepackage{graphicx}
\usepackage{latexsym}
\usepackage{epstopdf}

\usepackage{soul}
\biboptions{sort&compress}
\usepackage{setspace}
\usepackage{xcolor}

\newcommand{\bea}{\begin{eqnarray}}
\newcommand{\eea}{\end{eqnarray}}
\newcommand{\bes}{\begin{subequations}}
\newcommand{\ees}{\end{subequations}}

\newcommand{\ben}{\begin{equation}}
\newcommand{\een}{\end{equation}}

\newcommand{\sech}{\mbox{ sech}}
\usepackage[colorlinks=true,bookmarks=false,citecolor=blue,urlcolor=blue]{hyperref}

\usepackage{fancyhdr}
\makeatother

\begin{document} 
	\title{ Linear coupling effect induced beating non-degenerate vector solitons}
	
	\author[bdu1]{S. Stalin \corref{cor}}
	\author[bdu1]{M. Lakshmanan}
	\address[bdu1]{Department of Nonlinear Dynamics, School of Physics, Bharathidasan University, Tiruchirappalli - 620 024, Tamil Nadu, India} 
	
	\cortext[cor]{
		Email: stalin.cnld@gmail.com (S. Stalin)\newline 
		\indent \quad Email: lakshman.cnld@gmail.com (M. Lakshmanan)}
	\journal{}
	
	\setstretch{1.213}
	\begin{abstract}
	In this paper, we propose an alternative approach to generate a new class of beating vector solitons. Unlike earlier procedures that use dark-bright or bright-dark soliton solutions to generate beating solitons, the method described here utilizes non-degenerate vector soliton solutions of the Manakov system. It involves linear superposition of such soliton solutions along with an intensity switching mechanism facilitated by cross-coupling between the optical modes. We find that the obtained beating solitons collide elastically with themselves and keep their beating feature unchanged after the collision.	We also find that their beating nature can be controlled by allowing them to collide with degenerate beating solitons exhibiting energy-sharing collisions. The results presented in this work will provide new insights into beating solitons in Bose-Einstein condensates, nonlinear optics, and related areas of research.
		
		\noindent{\it Keywords:} Beating non-degenerate vector solitons, Beating degenerate vector solitons, Elastic and inelastic collisions of beating vector solitons. 
	\end{abstract}

	\maketitle
	

	\setstretch{1.3}
	\section{Introduction}
In physics, it is well known that when two waves have slightly different frequencies, their linear superposition produces a phenomenon called beats. This phenomenon arises from the interplay of constructive and destructive interference and is characterized by a beating frequency and a beating period. Recently, beating solitons (BSs), which are wave structures exhibiting beating behaviour, have been receiving much interest in soliton-supporting coupled nonlinear Schr\"odinger systems \cite{PGK-1,PGK-2,hoefer,PGK-3,li-chen-1}. A distinctive feature of such vector solitons is that they exhibit periodic oscillations in the density/intensity of individual components, while no beating effects occur in the total density or total intensity profiles. Based on the appearance of the total density profiles, BSs were classified as beating dark solitons and beating anti-dark solitons in the two-coupled system, and as beating bright soliton with a double-hump, beating dark soliton with a double valley, and beating bright solitons with a triple-hump in the three-coupled system \cite{PGK-3,li-chen-1}.

To generate this new class of vector solitons, different procedures have been adopted in the literature \cite{PGK-1,PGK-2,hoefer,li-chen-1,park,PGK-3,PGK-4,c-liu,wjche,cc-ding}. For example, beating effects of solitons in multi-component Bose-Einstein condensates (BECs) were induced  through a linear superposition of eigenstates (either dark-bright soliton pair or bright-dark soliton pair) in the effective quantum well \cite{li-chen-1}. Then, by utilizing the SU(2) and SU(3) rotation symmetries admitted by the coupled nonlinear Schr\"odinger (CNLS)  systems \cite{li-chen-1,PGK-1,PGK-2,PGK-3,PGK-4,park,c-liu} and the linear and/or nonlinear superposition of dark-bright vector solitons the BSs were generated \cite{c-liu,wjche}. The existence of this novel soliton states were observed experimentally in two-component BECs \cite{hoefer} and by numerical simulations in non-integrable systems \cite{PGK-3,PGK-4}. Very recently, multi-parameter vector BSs of different kinds have been reported in both the focusing and defocusing Manakov models, and their physical spectra have also been calculated analytically \cite{c-liu,wjche}. Along this line, it has been shown that two-component BECs with helicoidal spin-orbit coupling can give rise to beating stripe solitons \cite{cc-ding}. Further, it has been demonstrated that the presence of four-wave mixing effect can also induce beating effects. However, in this case, the total intensity profile exhibits spatio-temporal oscillations \cite{senwu}. We note here that the dynamics of beating solitons in the two-component BECs with Rabi coupling effect and the formation of asymmetric solitons have also been discussed in \cite{xl-li}. The beating soliton has also been reported in the coupled derivative NLS system \cite{wangxy} and the coupled Hirota equation \cite{lpan}.

From these studies, we infer that the generation of BSs has been achieved so far primarily through the superposition of dark-bright/bright-dark vector solitons. As pointed out in \cite{li-chen-1}, other vector solitons such as bright-bright and dark-dark vector solitons are not suitable for generating BSs, as they possess identical eigenvalues and identical density profiles.

Motivated by the above progress, in this work we propose an alternative approach to generate a novel type of BSs by considering self- and cross-coupling of orthogonally polarized modes, governed by the coupled partial differential equations,
\begin{eqnarray}
	iq_{1,z}+q_{1,tt}+2(|q_1|^2+|q_2|^2)q_1+\rho q_1+\nu q_2=0,\nonumber\\
	iq_{2,z}+q_{2,tt}+2(|q_1|^2+|q_2|^2)q_2-\rho q_2+\nu q_1=0,
	\label{eq1}
\end{eqnarray} 
as well as through the linear superposition of the nondegenerate vector solitons of the Manakov model \cite{manakov}
\begin{eqnarray}
	i\psi_{j,z}+\psi_{j, tt}+2(|\psi_1|^2+|\psi_2|^2)\psi_j=0,~j=1,2.\label{manakov}
\end{eqnarray}
This idea of using nondegenerate vector solitons to generate BSs is distinct from earlier works, where mixed bright-dark or dark-bright vector solitons were used to bring out BSs. The introduction of cross-coupling ($\nu$)  between the modes, along with self-coupling ($\rho$) among them, leads to an exchange of intensity or energy between the optical modes. Through this mechanism of intensity switching we are able to generate the desired beating solitons. In the above Eqs. (\ref{eq1}) and (\ref{manakov}), the dependent variables $q_j\equiv q_j(z,t)$'s and  $\psi_j\equiv \psi_j(z,t)$'s represent the complex field envelopes, and the suffices with respect to the independent variables $z$ and $t$ denote the partial derivatives with respect to those dimensionless variables. In nonlinear fiber optics, these independent variables generally represent the normalized distance along the fiber and retarded time, respectively. The model (\ref{eq1}) also appears in nonlinear optics for describing wave propagation in periodically twisted birefringent fibers \cite{potasek,radhakrishnan}, where the real constants $\nu$ and $\rho$ arise, respectively, from the periodic twist of the birefringent axes and the phase-velocity mismatch. Equation (\ref{eq1}) also arises in two-component BECs with linear Rabi coupling between the two population states \cite{qu,williams,xl-li}. 

The remaining part of the paper is organized as follows. In Section 2, we present the beating non-degenerate soliton solution of the CNLS system (\ref{eq1}) and analyse its associated beating features. Section 3 explores the collision dynamics of the obtained beating non-degenerate solitons and controlling of their beating nature through collision with beating degenerate solitons. Finally the results are summarized in Section 4. 
\section{Beating non-degenerate vector soliton}
As we mentioned earlier, to construct the beating vector soliton solutions for Eq. (\ref{eq1}) we map the nondegenerate vector soliton solutions of the Manakov system (\ref{manakov}) through the transformation \cite{potasek,radhakrishnan}
\begin{eqnarray}
	\begin{pmatrix}
		q_1\\
		q_2
	\end{pmatrix}=\begin{pmatrix}
		\cos\frac{\theta}{2}e^{i\Gamma z} & -\sin\frac{\theta}{2}e^{-i\Gamma z}\\
		\sin\frac{\theta}{2} e^{i\Gamma z} & \cos\frac{\theta}{2}e^{-i\Gamma z}
	\end{pmatrix}\begin{pmatrix}
		\psi_1\\
		\psi_2
	\end{pmatrix}.\label{trans}
\end{eqnarray}
In the above, $\Gamma=\sqrt{\rho^2+\nu^2}$,  $\theta=\tan^{-1}(\frac{\nu}{\rho})$, and $\psi_j$'s are the nondegenerate bright soliton solutions of the Manakov system (\ref{manakov}). Note that the above transformation (\ref{trans}) implies that Eq. (\ref{eq1}) is invariant under $\mathit{SU}(2)$ rotations and $|\cos\frac{\theta}{2}e^{i\Gamma z}|^2+|\sin\frac{\theta}{2}e^{i\Gamma z}|^2=1$.  
Using Eq. (\ref{trans}),  $\mathit{SU}(2)$ rotated beating non-degenerate one soliton solution of Eq. (\ref{eq1}) is obtained as
\begin{eqnarray}
q_1(z,t)=	\cos\frac{\theta}{2}e^{i\Gamma z}\psi_1-\sin\frac{\theta}{2} e^{-i\Gamma z}\psi_2,\nonumber\\
q_2(z,t)=	\sin\frac{\theta}{2}e^{i\Gamma z}\psi_1+\cos\frac{\theta}{2}e^{-i\Gamma z}\psi_2, \label{eq4}
\end{eqnarray}
where $\psi_j$'s are the non-degenerate one soliton solution of the Manakov system \cite{stalin1,stalin2,stalin-review}, which is of the form
\begin{eqnarray}
&&\hspace{-0.5cm}\psi_1=\frac{1}{D}[c_{11}e^{i\eta_{1I}}\cosh(\xi_{1R}+\phi_1)]\nonumber\\
&&\hspace{-0.5cm}\psi_2=\frac{1}{D}[c_{21}e^{i\xi_{1I}}\cosh(\eta_{1R}+\phi_2)]\nonumber\\
&&\hspace{-0.5cm}D=[c_{12}\cosh(\eta_{1R}+\xi_{1R}+\phi_1+\phi_2+b_1)\nonumber\\
&&\hspace{0.7cm}+c_{13}\cosh(\eta_{1R}-\xi_{1R}+\phi_2-\phi_1+b_2)], \label{5}
\end{eqnarray}
where 
\begin{eqnarray}
&&\hspace{-1.3cm}\eta_{1R}=k_{1R}(t-2k_{1I}z), ~~\xi_{1R}=l_{1R}(t-2l_{1I}z), \nonumber\\
&&\hspace{-1.3cm}\eta_{1I}=k_{1I}t+(k_{1R}^2-k_{1I}^2)z,~ \xi_{1I}=l_{1I}t+(l_{1R}^2-l_{1I}^2)z,\nonumber\\
&&\hspace{-1.3cm}c_{11}=e^{\frac{\Delta_{11}+\rho_1}{2}}, ~c_{21}=e^{\frac{\Delta_{12}+\rho_2}{2}}, ~c_{12}=e^{\frac{R_3}{2}},  ~c_{13}=e^{\frac{R_1+R_2}{2}},\nonumber\\ 
&&\hspace{-1.3cm}\phi_1=\frac{\Delta_{11}-\rho_1}{2}, ~\phi_2=\frac{\Delta_{12}-\rho_2}{2},~~e^{\rho_j}=\alpha_1^{(j)}, ~j=1,2,
\nonumber\\ 
&&\hspace{-1.3cm}b_1=\frac{1}{2}\ln\frac{(k_1^*-l_1^*)}{(l_1-k_1)},~b_2=\frac{1}{2}\ln\frac{(k_1^*+l_1)(k_1-l_1)}{(k_1+l_1^*)(l_1-k_1)},\nonumber\\
&&\hspace{-1.3cm}e^{R_{1}}=\frac{|\alpha_{1}^{(1)}|^2}{(k_{1}+k_{1}^{*})^{2}}, ~e^{R_{2}}=\frac{ |\alpha_{1}^{(2)}|^2}{(l_{1}+l_{1}^*)^{2}}, ~e^{R_{3}}=\frac{|k_{1}-l_{1}|^{2}}{|k_{1}+l_{1}^*|^2}e^{R_1+R_2}, \nonumber\\
 &&\hspace{-1.3cm}e^{\Delta_{11}}=\frac{(k_{1}-l_{1})\alpha_{1}^{(1)}|\alpha_{1}^{(2)}|^2}{(k_{1}+l_{1}^*)(l_{1}+l_{1}^*)^{2}}, ~\text{and}~
e^{\Delta_{12}}=\frac{(l_{1}-k_{1})|\alpha_{1}^{(1)}|^2\alpha_{1}^{(2)}}{(k_{1}+k_{1}^*)^{2}(k_{1}^*+l_{1})}.~~~~~
\end{eqnarray}
In the above and following expressions, the subscripts $R$ and $I$ represent the real and imaginary parts of the respective variables or constants. The beating non-degenerate one soliton solution (\ref{eq4}) is characterized by four arbitrary complex parameters $k_1$, $l_1$, and $\alpha_1^{(j)}$, $j=1,2$, and in addtion two real system parameters $\rho$ and $\nu$. 
As it was shown by the present authors and their collaborators in Refs. \cite{stalin1,stalin2,stalin-review} and illustrated below in Fig. \ref{fig2}, the non-degenerate vector soliton, in the absence of self and cross-coupling terms,  admits a double-hump, a flat-top and a standard single-hump intensity profiles for $k_{1I}=l_{1I}$.  However, their linear superposition, through Eq. (\ref{trans}), along with cross and self coupling of optical modes give rise to the beating effects in the structures of non-degenerate solitons.       

\begin{figure*}[ht]
	\centering
	\includegraphics[width=0.7\linewidth]{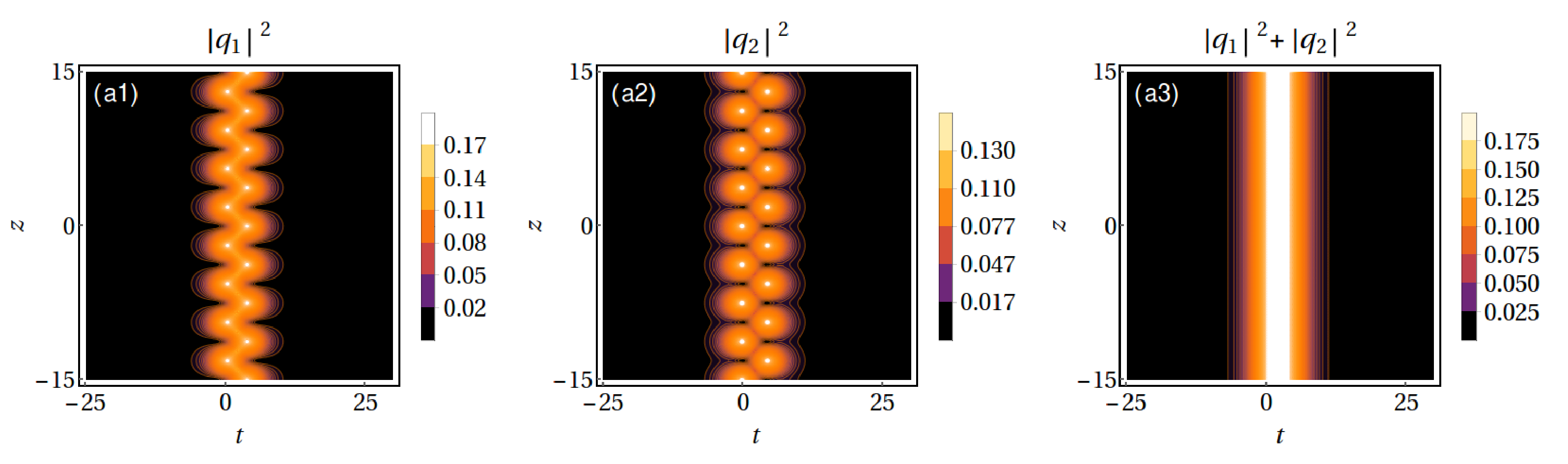}\\
	\includegraphics[width=0.7\linewidth]{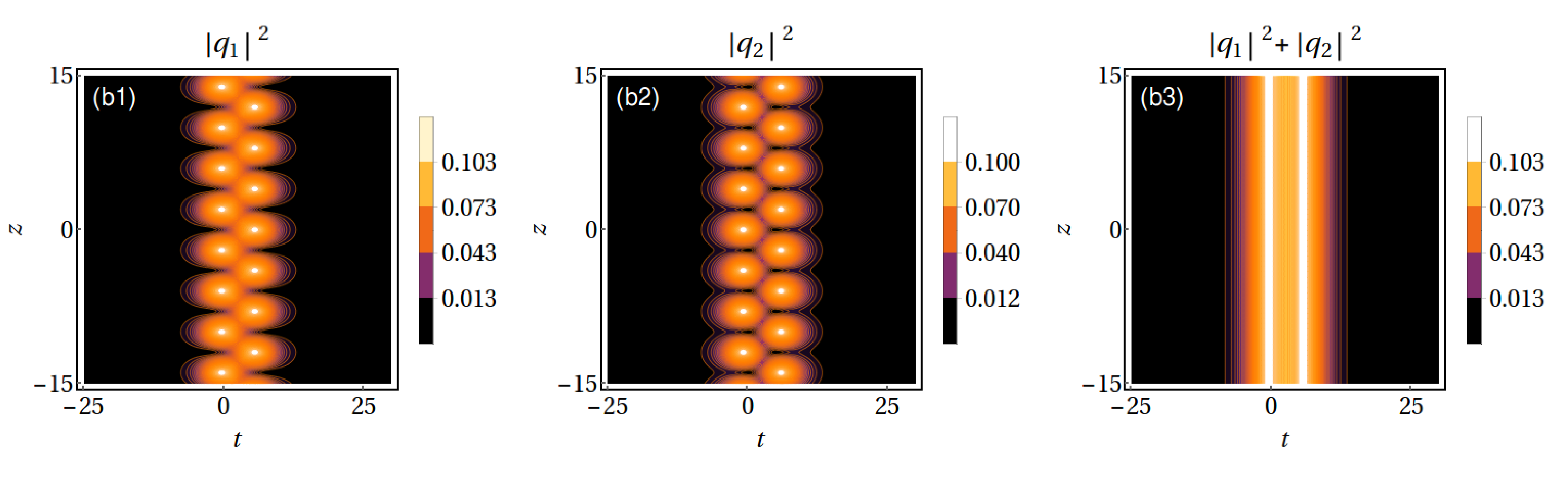}\\
	\includegraphics[width=0.7\linewidth]{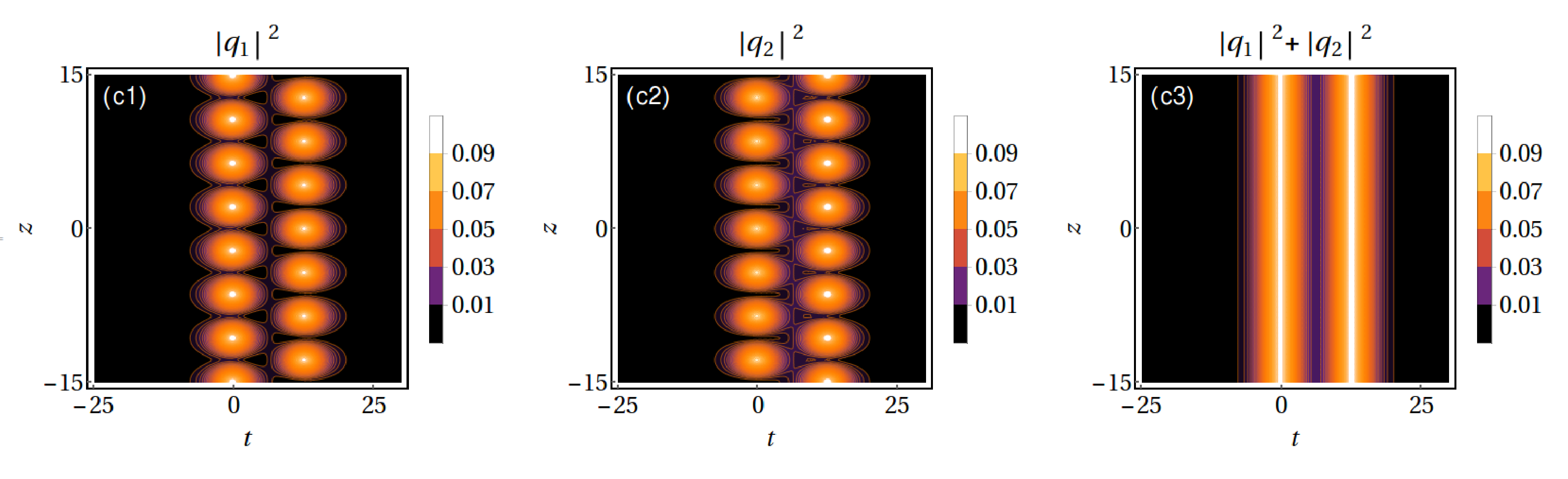}\vspace{-0.25cm}
	\caption{Self ($\rho=0.25$) and cross ($\nu=0.7$) coupling induced beating non-degenerate fundamental soliton. Panels (a1)-(a3): $k_1=0.565$, $l_1=0.3$, $\alpha_1^{(1)}=0.44+0.51i$, and $\alpha_1^{(2)}=0.43+0.5i$. Panels (b1)-(b3): $k_1=0.425$, $l_1=0.3$, $\alpha_1^{(1)}=0.44+0.51i$, and $\alpha_1^{(2)}=0.43+0.5i$. Panels (c1)-(c3): $k_1=0.32$, $l_1=0.34$, $\alpha_1^{(1)}=0.55$, and $\alpha_1^{(2)}=0.45$. }
	\label{fig1}
\end{figure*}

To understand the beating nature and generation mechanism of beating nondegenerate solitons, we derive the intensity expressions from the corresponding analytical forms (\ref{eq4}) of the modes $q_1$ and $q_2$ with $k_{1I}=l_{1I}$. The intensities are found to be
\begin{subequations}
\begin{eqnarray}
	&&\hspace{-1cm}|q_1|^2=P_1\big(\cos^2\frac{\theta}{2}|c_{11}|^2P_2+\sin^2\frac{\theta}{2}|c_{21}|^2P_2^{-1}\nonumber\\
	&&\hspace{-0.1cm}-\sin\theta e^{-iQ}c_{11}c_{21}^*\cos[(2\Gamma+ k_{1R}^2-l_{1R}^2)z+Q]\big),~~~\label{6a}\\
	&&\hspace{-1cm}|q_2|^2=P_1\big(\sin^2\frac{\theta}{2}|c_{11}|^2P_2+\cos^2\frac{\theta}{2}|c_{21}|^2P_2^{-1}\nonumber\\
	&&\hspace{-0.1cm}+\sin\theta e^{-iQ}c_{11}c_{21}^*\cos[(2\Gamma+k_{1R}^2-l_{1R}^2)z+Q]\big),~~~\label{6b}
\end{eqnarray}
where \begin{eqnarray}
&&\hspace{-1.1cm}P_1=\frac{1}{D_1^2}(\cosh(\xi_{1R}+\phi_1)\cosh(\eta_{1R}+\phi_2)),\nonumber\\
&&\hspace{-1.1cm}D_1=[c_{12}\cosh(\eta_{1R}+\xi_{1R}+\phi_1+\phi_2+b_1)\nonumber\\
&&\hspace{0.1cm}+c_{13}\cosh(\eta_{1R}-\xi_{1R}+\phi_2-\phi_1+b_1)], \nonumber\\
&&\hspace{-1.1cm}P_2=\frac{\cosh(\xi_{1R}+\phi_1)}{\cosh(\eta_{1R}+\phi_2)},~|c_{11}|^2=\frac{(k_{1R}-l_{1R})|\alpha_1^{(1)}|^2|\alpha_1^{(2)}|^2}{4l_{1R}^2(k_{1R}+l_{1R})}, \nonumber\\
&&\hspace{-1.1cm}|c_{21}|^2=\frac{(l_{1R}-k_{1R})|\alpha_1^{(1)}|^2|\alpha_1^{(2)}|^2}{4k_{1R}^2(k_{1R}+l_{1R})},~ c_{13}=\frac{|\alpha_1^{(1)}||\alpha_1^{(2)}|}{4k_{1R}l_{1R}},\nonumber\\ 
&&\hspace{-1.1cm}\phi_1=\frac{1}{2}\ln \frac{(k_{1R}-l_{1R})|\alpha_1^{(2)}|^2}{4l_{1R}^2(k_{1R}+l_{1R})},~\phi_2=\frac{1}{2}\ln \frac{(l_{1R}-k_{1R})|\alpha_1^{(1)}|^2}{4k_{1R}^2(k_{1R}+l_{1R})},\nonumber\\
&&\hspace{-1.1cm}c_{12}=\frac{(k_{1R}-l_{1R})|\alpha_1^{(1)}||\alpha_1^{(2)}|}{4k_{1R}l_{1R}(k_{1R}+l_{1R})},~ Q=\frac{1}{2}\tan^{-1}\big(\frac{2AB}{A^2-B^2}\big),\nonumber\\
&&\hspace{-1.1cm}b_1=\frac{1}{2}\ln\frac{(k_{1R}-l_{1R})}{(l_{1R}-k_{1R})},~ A=(\alpha_{1R}^{(1)}\alpha_{1R}^{(2)}+\alpha_{1I}^{(1)}\alpha_{1I}^{(2)}),\nonumber\\ &&\hspace{-1.1cm}B=(\alpha_{1I}^{(1)}\alpha_{1R}^{(2)}-\alpha_{1I}^{(2)}\alpha_{1R}^{(1)}),~
\eta_{1R}=k_{1R}(t-2k_{1I}z), \nonumber\\
&&\hspace{-1.1cm}\text{and}~~\xi_{1R}=l_{1R}(t-2k_{1I}z). 	\label{6c}
\end{eqnarray}
\end{subequations}
It is evident from expressions (\ref{6a})-(\ref{6c}) and from Fig. \ref{fig1}, the spatially periodic oscillations along the propagation direction $z$ in the intensities of the individual components are induced by two main factors. The presence of $\Gamma$ in the oscillatory term
 $\cos((2\Gamma+ k_{1R}^2-l_{1R}^2)z+Q)$ implies that the linear self- and cross-coupling induce oscillations through intensity switching between the modes $q_1$ and $q_2$. On the other hand, the appearance of the real parts of the wave numbers $k_1$ and $l_1$ in the oscillatory term, arising from the linear superposition of the non-degenerate solitons, also contributes to the periodic oscillations along the $z$-direction. These oscillations are characterized by the beating frequency: $\omega=|2\Gamma+k_{1R}^2-l_{1R}^2|$ and the corresponding spatial period: $Z=\frac{2\pi}{|2\Gamma+k_{1R}^2-l_{1R}^2|}$. The amplitude of oscillations is given by $\sin\theta |c_{11}||c_{21}|$. The beating nondegenerate soliton profiles corresponding to the three distinct nondegenerate soliton profiles of the Manakov system (\ref{manakov}), shown in Figs. \ref{fig2}(a1)-(a2), \ref{fig2}(b1)-(b2), and \ref{fig2}(c1)-(c2), are displayed in Figs. \ref{fig1}(a1)-(a3), \ref{fig1}(b1)-(b3), and \ref{fig1}(c1)-(c3), respectively. If $\rho=\nu=0$, there is no exchange of intensity between the components. As a result, no oscillation occurs in the individual modes.

In Ref. \cite{li-chen-1}, the term beating solitons is used based on the structure of the total intensity profiles. In a similar way, we compute the total intensity here using Eqs. (\ref{6a}) and (\ref{6b}) to substantiate our naming of the above profiles as BS. By doing so, the total intensity is calculated as
\begin{equation}
	|q_1|^2+|q_2|^2=P_1\big(|c_{11}|^2P_2+|c_{21}|^2P_2^{-1}\big). \label{7}
\end{equation}
Here, $P_1$, $P_2$, $c_{11}$, and $c_{21}$ are defined as above. From Eq. (\ref{7}), we observe that the total intensity does not contain any oscillatory terms, and hence, no beating behavior appears in the total intensity profile. Rather, the total intensity profile exhibits a double-hump structure, as confirmed by Eq. (\ref{7}) and the last column of Fig. \ref{fig1} (see Figs. \ref{fig1}(a3), \ref{fig1}(b3), and \ref{fig1}(c3)). Therefore, the beating soliton presented in Eq. (\ref{eq4}) is referred to as a beating soliton with a double-hump structure. The double-hump soliton appearing in the total intensity profile also propagates with the same group velocity ($v_g=2k_{1I}$) as the BS in the individual components. 
\begin{figure*}
	\centering	
	\includegraphics[width=0.45\linewidth]{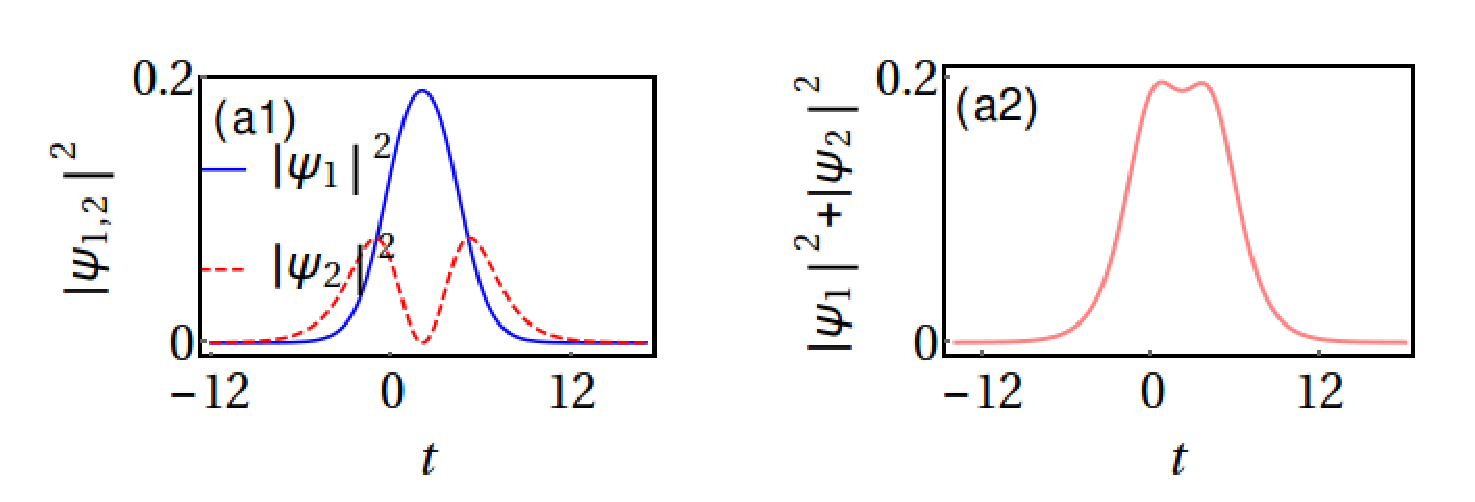}\\
	\includegraphics[width=0.45\linewidth]{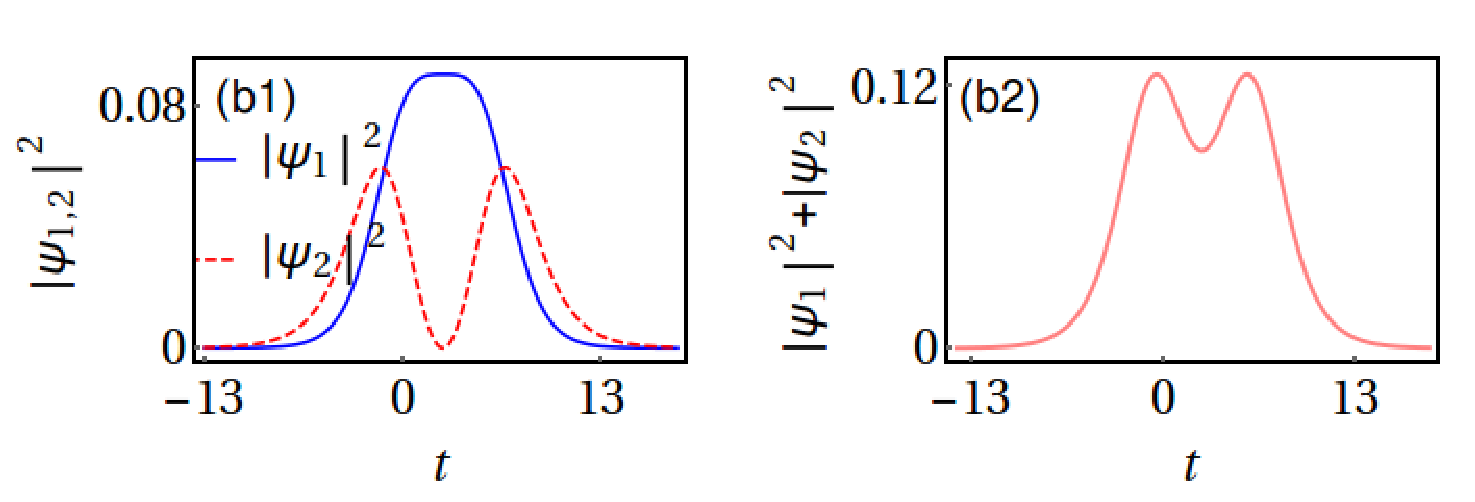}\\
	\includegraphics[width=0.45\linewidth]{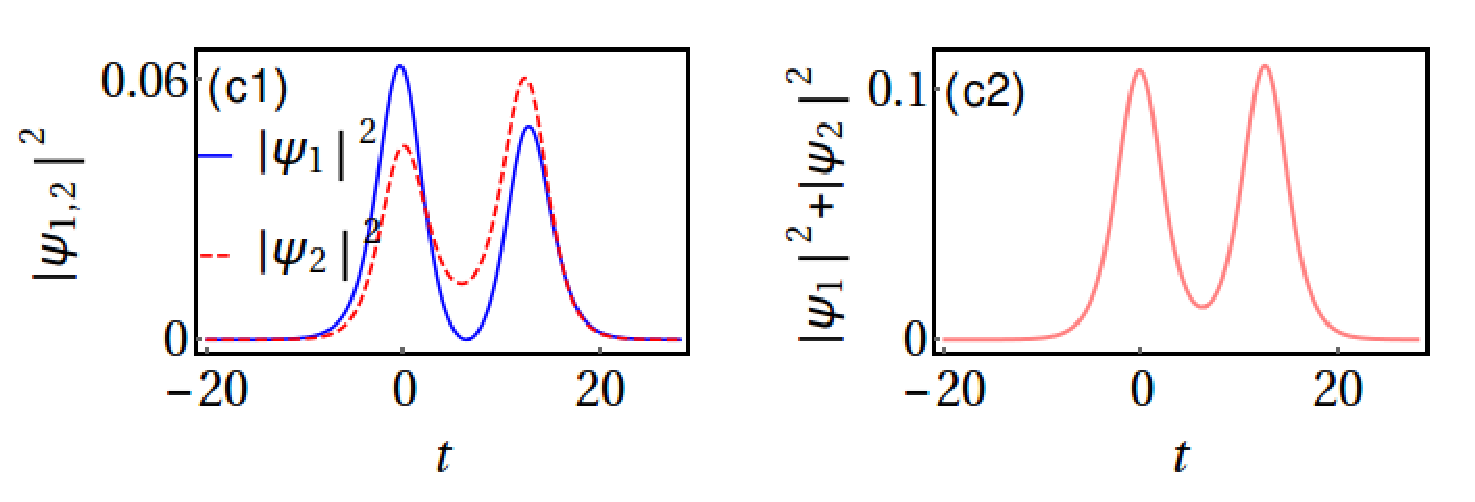}\vspace{-0.25cm}
	\caption{Non-degenerate fundamental soliton of the Manakov system in the absence of self ($\rho$) and cross ($\nu$) coupling effects. Panels (a1) and (a2): $k_1=0.565$, $l_1=0.3$, $\alpha_1^{(1)}=0.44+0.51i$, and $\alpha_1^{(2)}=0.43+0.5i$. Panels (b1) and (b2): $k_1=0.425$, $l_1=0.3$, $\alpha_1^{(1)}=0.44+0.51i$, and $\alpha_1^{(2)}=0.43+0.5i$. Panels (c1) and (c2): $k_1=0.32$, $l_1=0.34$, $\alpha_1^{(1)}=0.55$, and $\alpha_1^{(2)}=0.45$. }
	\label{fig2}
\end{figure*}

\begin{figure*}
	\centering	
	\includegraphics[width=0.5\linewidth]{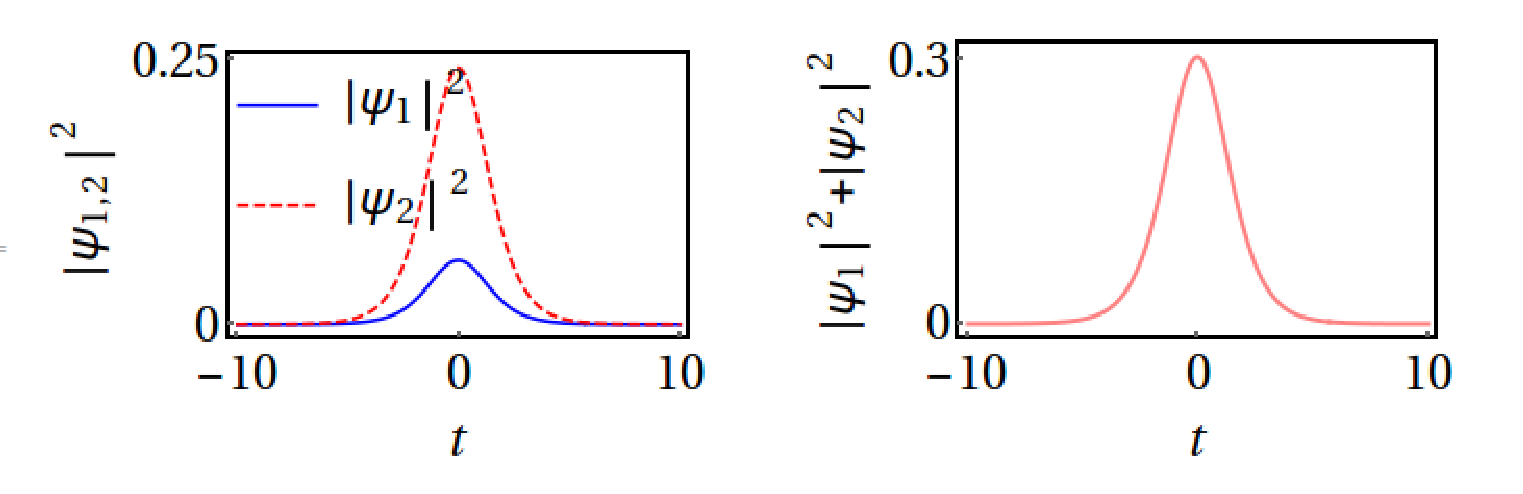}\vspace{-0.25cm}
	\caption{A stationary degenerate vector bright soliton of the Manakov system in the absence of self and cross-coupling effect is shown here. The parameter values are $k_1=l_1=0.55$, $\alpha_1^{(1)}=0.5$,  $\alpha_1^{(2)}=1$, and $\rho=\nu=0$.  }
	\label{fig3}
\end{figure*}
\begin{figure*}
	\centering
	\includegraphics[width=1.0\linewidth]{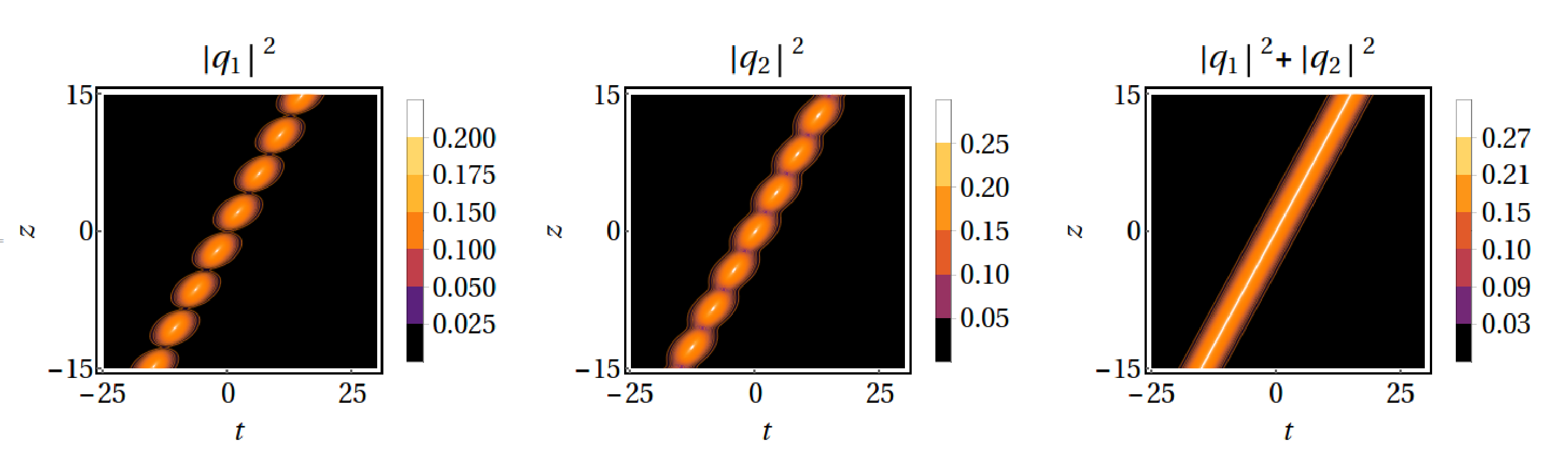}\vspace{-0.25cm}
	\caption{Beating-degenerate vector soliton: $\rho=0.25$, $\nu=0.7$, $k_1=l_1=0.55+0.5i$, $\alpha_1^{(1)}=0.5$, and $\alpha_1^{(2)}=1$. }
	\label{fig4}
\end{figure*}

We remark that the spatial oscillations in the BS can be suppressed by setting either $c_{11}$  or $c_{21}$ (or both) to zero. This effect can be achieved by assigning zero values to the soliton parameters $\alpha_1^{(j)}$'s. That is, if $\alpha_1^{(1)}=0$ and $\alpha_1^{(2)}\neq 0$, or $\alpha_1^{(1)}\neq 0$ and $\alpha_1^{(2)}=0$, then as a result, we obtain either $\psi_1=0$, and $\psi_2=\frac{\alpha_1^{(2)}e^{\xi_1}}{1+e^{\xi_1+\xi_1^*+R_2}}$, or $\psi_1=\frac{\alpha_1^{(1)}e^{\eta_1}}{1+e^{\eta_1+\eta_1^*+R_1}}$, $\psi_2=0$, respectively, from Eq. (\ref{5}). Further, when we impose $l_{1}=l_{1R}+il_{1I}=k_{1}=k_{1R}+ik_{1I}$, the limiting vector bright soliton can be obtained \cite{stalin1,stalin2}, which is same as the form (\ref{eq4})  but with $\psi_j=\alpha_1^{(j)}e^{\eta_1}/(1+e^{\eta_1+\eta_1^*+R})$, $j=1,2$,  $\eta_1=k_1t+ik_1^2z$, $e^{R}=(|\alpha_1^{(1)}|^2+|\alpha_1^{(2)}|^2)/(k_1+k_1^*)^2$. This vector bright soliton depicted in Fig. \ref{fig3} for $\rho=\nu=0$ cannot be used to construct a BS solution since it is considered as a degenerate soliton due to the presence of identical eigenvalues (or wave numbers) in both the components \cite{li-chen-1,stalin1,stalin2}. However, the formalism presented in this work can induce the beating effects in such degenerate vector solitons by introducing linear couplings between the modes \cite{potasek,radhakrishnan}.
This beating phenomenon can be visualized from Fig. \ref{fig4}, by imposing the constraint $k_{1}=l_{1}$ in the solution (\ref{eq4}). We note here that efforts have been made to understand the beating effects of non-degenerate solitons through the eigenstates in quantum mechanics \cite{li-chen-1,Qin1}. 
    
\begin{figure*}
	\centering
	\includegraphics[width=1.0\linewidth]{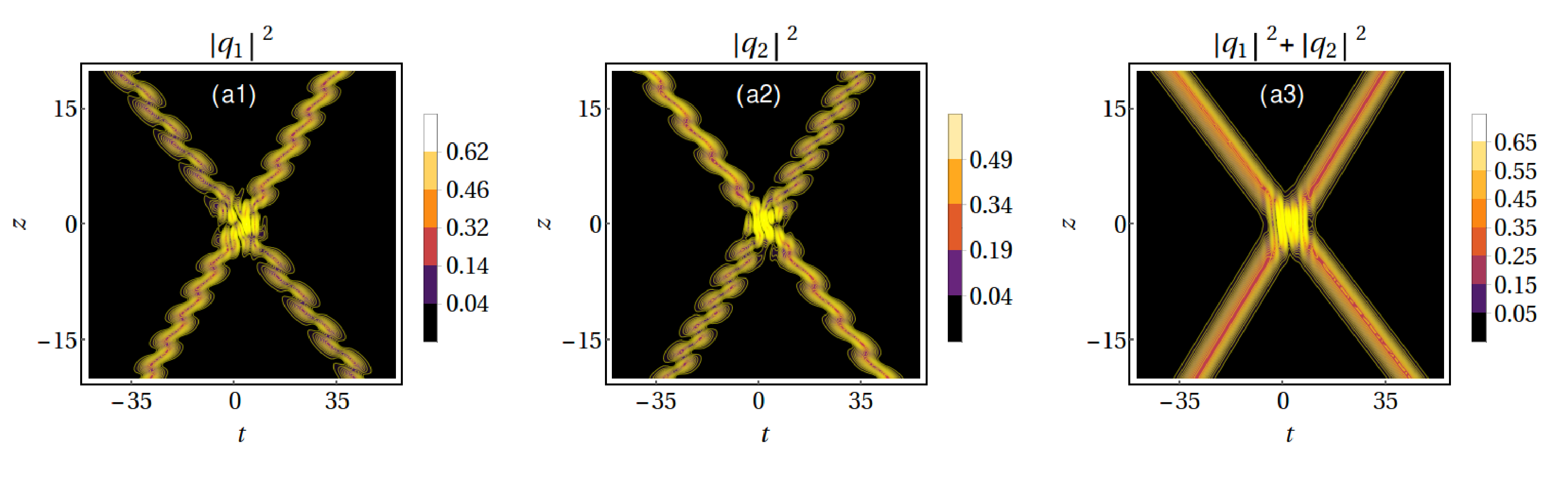}\\
	\includegraphics[width=1.0\linewidth]{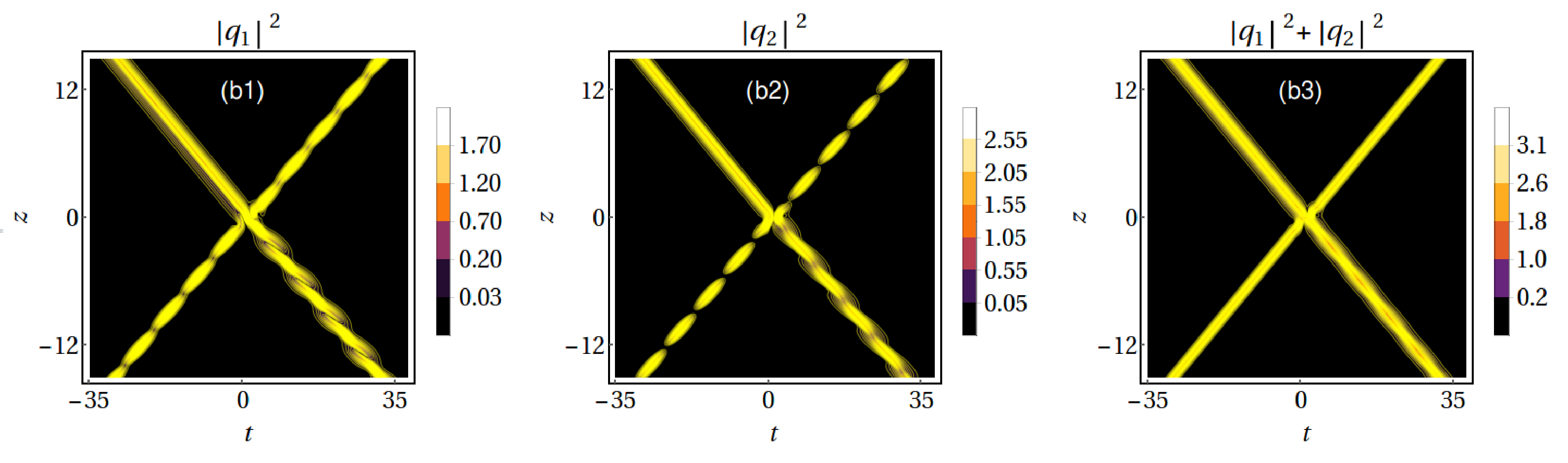}
	\caption{While panels (a1)-(a3) illustrate the elastic collision dynamics of two beating non-degenerate vector solitons, panels (b1)-(b3) depict how the beating effects of a non-degenerate soliton can be controlled through collision with a degenerate beating soliton. Panels (a1)-(a3): $\rho=0.25$, $\nu=0.8$, $k_1=0.565+0.8i$, $l_1=0.3+0.8i$, $k_2=0.3-i$, $l_2=0.57-i$,  $\alpha_1^{(1)}=0.44+0.51i$, $\alpha_1^{(2)}=0.43+0.5i$, $\alpha_2^{(1)}=0.45+0.45i$, $\alpha_2^{(2)}=0.55$. Panels (b1)-(b3): $\rho=0.25$, $\nu=0.9$, $k_1=l_1=1.3+i$, $k_2=0.6-i$, $l_2=1.5-i$,  $\alpha_1^{(1)}=0.5+0.5i$, $\alpha_1^{(2)}=0.5$, $\alpha_2^{(1)}=1$, $\alpha_2^{(2)}=0.45+0.6i$. 
	}
	\label{fig5}
\end{figure*}

\section{Collision dynamics of beating vector solitons}
Several natural questions will arise regarding the interaction of beating nondegenerate solitons with themselves and with other soliton types, the possibility of controlling their beating effects through collisions, and the influence of self- and cross-coupling terms on the resulting collision dynamics. To answer these questions, we consider the following two situations: (i) collision between two beating non-degenerate solitons, and (ii) collision between a beating degenerate soliton and a beating non-degenerate soliton. We present below the asymptotic analysis corresponding to the former case only, while for the latter case a similar analysis can be carried out. 
\subsection{Asymptotic analysis} To analyse the interaction between beating non-degenerate solitons, we consider the choice, $k_{jI}=l_{jI}$, $k_{jR},~l_{jR}>0$, $j=1,2$, and $k_{1I}>k_{2I}$ in the wave variables $\eta_{jR}=k_{jR}(t-2k_{jI}z)$, $\xi_{jR}=l_{jR}(t-2l_{jI}z)$, $j=1,2$, of the beating non-degenerate two-soliton solution (Note: To obtain the multi-beating non-degenerate soliton solutions, one has to map the non-degenerate multi-soliton solutions of the Manakov system (\ref{manakov}). The exact form of which are given in our earlier papers \cite{stalin1} and \cite{stalin2}). We also consider the asymptotic limit $z\rightarrow \pm \infty$ to deduce the following expressions for BSs 1 and 2 before and after the collision. These asymptotic expressions of BSs are given below:   

{ \bf Beating soliton 1: } $\eta_{1R}, ~\xi_{1R}\approx 0$,  $\eta_{2R}, ~\xi_{2R}\rightarrow \mp\infty$ as $z\rightarrow \mp\infty$\\
The asymptotic forms of BS 1 before and after the collision are identified as follows.
 \bes\begin{eqnarray}
	&&q_1^{1\mp}=\cos\frac{\theta}{2}e^{i\Gamma z}\psi_1^{1\mp}-\sin\frac{\theta}{2}e^{-i\Gamma z}\psi_2^{1\mp},\\
	&&q_2^{1\mp}=\sin\frac{\theta}{2}e^{i\Gamma z}\psi_1^{1\mp}+\cos\frac{\theta}{2}e^{-i\Gamma z}\psi_2^{1\mp},
\end{eqnarray}
where \begin{eqnarray}
&&\hspace{-1.1cm}\psi_1^{1\mp}=\frac{2k_{1R}}{P_1^\mp}e^{i(\theta_1^{1\mp}+\eta_{1I})}\cosh(\xi_{1R}+\phi_1^{1\mp}),\nonumber\\
&&\hspace{-1.1cm}\psi_2^{1\mp}=\frac{2l_{1R}}{P_1^\mp}e^{i(\theta_2^{1\mp}+\xi_{1I}+\frac{\pi}{2})}\cosh(\eta_{1R}+\phi_2^{1\mp}),\nonumber\\
&&\hspace{-1.1cm}P_1^\mp=\hat{c}_{11}\cosh(\eta_{1R}+\xi_{1R}+\phi_1^{1\mp}+\phi_2^{1\mp}+b_1)\nonumber\\
&&\hspace{0.2cm}+\hat{c}_{11}^{-1}\cosh(\eta_{1R}-\xi_{1R}+\phi_2^{1\mp}-\phi_1^{1\mp}+b_1), \nonumber\\
&&\hspace{-1.1cm}\hat{c}_{11}=(\frac{k_{1R}-l_{1R}}{k_{1R}+l_{1R}})^{\frac{1}{2}},~\phi_1^{1-}=\frac{1}{2}\ln\frac{(k_{1R}-l_{1R})|\alpha_1^{(2)}|^2}{4l_{1R}^2(k_{1R}+l_{1R})},\nonumber\\
&&\hspace{-1.1cm} \phi_2^{1-}=\frac{1}{2}\ln\frac{(l_{1R}-k_{1R})|\alpha_1^{(1)}|^2}{4k_{1R}^2(k_{1R}+l_{1R})}, \nonumber\\
&&\hspace{-1.1cm}\phi_1^{1+}=\frac{1}{2}\log\frac{(k_{1R}-l_{1R})(k_{2R}-l_{1R})^2(l_{1R}-l_{2R})^4|\alpha_1^{(2)}|^2}{4l_{1R}^2(k_{1R}+l_{1R})(k_{2R}+l_{1R})^2(l_{1R}+l_{2R})^4},\nonumber\\
&&\hspace{-1.1cm}\phi_2^{1+}=\frac{1}{2}\log\frac{(l_{1R}-k_{1R})(k_{1R}-l_{2R})^2(k_{1R}-k_{2R})^4|\alpha_1^{(1)}|^2}{4k_{1R}^2(k_{1R}+l_{1R})(k_{1R}+l_{2R})^2(k_{1R}+k_{2R})^4}. ~~
\end{eqnarray}\ees
	Here, the superscript $1\mp$ denotes BS 1 before ($-$) and after ($+$) the collision, while the subscripts 1 and 2 indicate the mode numbers.\\
{ \bf Beating soliton 2: } $\eta_{2R}, \xi_{2R}\approx 0$,  $\eta_{1R}, \xi_{1R}\rightarrow \pm\infty$ as $z\rightarrow \mp\infty$\\ 
The expressions of BS 2 before and after the collision are derived as follows.
\bes\begin{eqnarray}
	&&q_1^{2\mp}=\cos\frac{\theta}{2}e^{i\Gamma z}\psi_1^{2\mp}-\sin\frac{\theta}{2}e^{-i\Gamma z}\psi_2^{2\mp},\\
	&&q_2^{2\mp}=\sin\frac{\theta}{2}e^{i\Gamma z}\psi_1^{2\mp}+\cos\frac{\theta}{2}e^{-i\Gamma z}\psi_2^{2\mp},
\end{eqnarray}
where\begin{eqnarray}
&&\hspace{-1.4cm}	\psi_1^{2\mp}=\frac{2k_{2R}}{P_2^\mp}e^{i(\theta_1^{2\mp}+\eta_{2I})}\cosh(\xi_{2R}+\phi_1^{2\mp}),\nonumber\\
&&\hspace{-1.4cm} \psi_2^{2\mp}=\frac{2l_{2R}}{P_2^\mp}e^{i(\theta_2^{2\mp}+\xi_{2I})}\cosh(\eta_{2R}+\phi_2^{2\mp}), \nonumber\\
&&\hspace{-1.4cm}P_2^\mp=\hat{c}_{12}\cosh(\eta_{2R}+\xi_{2R}+\phi_1^{2\mp}+\phi_2^{2\mp}+b_2)\nonumber\\
&&\hspace{0.25cm}+\hat{c}_{12}^{-1}\cosh(\eta_{2R}-\xi_{2R}+\phi_2^{2\mp}-\phi_1^{2\mp}+b_2), \nonumber\\
&&\hspace{-1.4cm} \hat{c}_{12}=(\frac{k_{2R}-l_{2R}}{k_{2R}+l_{2R}})^{\frac{1}{2}},~b_2=\frac{1}{2}\ln\frac{(k_{2R}-l_{2R})}{(l_{2R}-k_{2R})}, \nonumber\\
&&\hspace{-1.4cm}\phi_1^{2-}=\frac{1}{2}\ln\frac{(k_{2R}-l_{2R})(k_{1R}-l_{2R})^2(l_{1R}-l_{2R})^4|\alpha_2^{(2)}|^2}{4l_{2R}^2(k_{2R}+l_{2R})(k_{1R}+l_{2R})^2(l_{1R}+l_{2R})^4},\nonumber\\
&&\hspace{-1.4cm} \phi_2^{2-}=\frac{1}{2}\ln\frac{(l_{2R}-k_{2R})(k_{2R}-l_{1R})^2(k_{1R}-k_{2R})^4|\alpha_2^{(1)}|^2}{4k_{2R}^2(k_{2R}+l_{2R})(k_{2R}+l_{1R})^2(k_{1R}+k_{2R})^4},\nonumber\\
&&\hspace{-1.4cm} \phi_1^{2+}=\frac{1}{2}\ln\frac{(k_{2R}-l_{2R})|\alpha_2^{(2)}|^2}{4l_{2R}^2(k_{2R}+l_{2R})},~\phi_2^{2+}=\frac{1}{2}\ln\frac{(l_{2R}-k_{2R})|\alpha_2^{(1)}|^2}{4k_{2R}^2(k_{2R}+l_{2R})}.~~~~~~~~~
\end{eqnarray}\ees
Here also, the superscript $2\mp$ denotes BS 2 before ($-$) and after ($+$) the collision, while the subscripts 1 and 2 indicate the mode numbers. From the above asymptotic expressions, it is clear that the phase terms before and after interaction are related by
\bes\begin{equation}
	\phi_j^{1+}=\phi_j^{1-}+\varphi_j,~	\phi_j^{2+}=\phi_j^{2-}-\varphi_{j+2},~j=1,2,
\end{equation}
where \begin{eqnarray}
&&\varphi_1=\frac{1}{2}\ln\frac{(k_{2R}-l_{1R})^2(l_{1R}-l_{2R})^4}{(k_{2R}+l_{1R})^2(l_{1R}+l_{2R})^4},\nonumber\\ &&\varphi_2=\frac{1}{2}\ln\frac{(k_{1R}-l_{2R})^2(k_{1R}-k_{2R})^4}{(k_{1R}+l_{2R})^2(k_{1R}+k_{2R})^4},\nonumber\\ &&\varphi_3=\frac{1}{2}\ln\frac{(k_{1R}-l_{2R})^2(l_{1R}-l_{2R})^4}{(k_{1R}+l_{2R})^2(l_{1R}+l_{2R})^4}, \nonumber\\ \text{and}~~&&\varphi_4=\frac{1}{2}\ln\frac{(k_{2R}-l_{1R})^2(k_{1R}-k_{2R})^4}{(k_{2R}+l_{1R})^2(k_{1R}+k_{2R})^4}. 
\end{eqnarray}\ees
The above detailed asymptotic analysis as well as from Fig. \ref{fig5}(a1)-(a3), one can observe that the beating non-degenerate solitons undergo elastic collision as in the case of non-degenerate solitons of the Manakov system in the absence of linear couplings \cite{stalin1,stalin2}. However, to understand the role of the linear coupling parameters in this elastic collision, we evaluate the intensities of the colliding BSs in the asymptotic limits $z\rightarrow \mp\infty$.  The intensity expression corresponding to beating soliton $1$ is given by
\begin{subequations}
\begin{eqnarray}
	&&\hspace{-1.5cm}|q_1^{1\mp}|^2=4P_{11}^{\mp}\bigg(k_{1R}^2\cos^2\frac{\theta}{2}P_{21}^{\mp}+l_{1R}^2\sin^2\frac{\theta}{2}(P_{21}^{\mp})^{-1}\label{11a}\\
	&&\hspace{-0.7cm}-k_{1R}l_{1R}\sin\theta\cos[(2\Gamma+k_{1R}^2-l_{1R}^2)z+\theta_1^{1\mp}-\theta_2^{1\mp}-\frac{\pi}{2}]\bigg),\nonumber\\
	&&\hspace{-1.5cm}|q_2^{1\mp}|^2=4P_{11}^{\mp}\bigg(k_{1R}^2\sin^2\frac{\theta}{2}P_{21}^{\mp}+l_{1R}^2\cos^2\frac{\theta}{2}(P_{21}^{\mp})^{-1}\label{11b}\\
	&&\hspace{-0.7cm}+k_{1R}l_{1R}\sin\theta\cos[(2\Gamma+k_{1R}^2-l_{1R}^2)z+\theta_1^{1\mp}-\theta_2^{1\mp}-\frac{\pi}{2}]\bigg),\nonumber
\end{eqnarray}
where \begin{eqnarray}
&&\hspace{-1.0cm}P_{11}^{\mp}=\frac{1}{D_{11}^{\mp2}}(\cosh(\xi_{1R}+\phi_1^{1\mp})\cosh(\eta_{1R}+\phi_2^{1\mp})),\nonumber\\ &&\hspace{-1.0cm}D_{11}^{\mp}=[\hat{c}_{11}\cosh(\eta_{1R}+\xi_{1R}+\phi_1^{1\mp}+\phi_2^{1\mp}+b_1)\nonumber\\
&&\hspace{0.25cm}+\hat{c}_{11}^{-1}\cosh(\eta_{1R}-\xi_{1R}+\phi_2^{1\mp}-\phi_1^{1\mp}+b_1)], \nonumber\\
&&\hspace{-1.0cm} \text{and}~~P_{21}^{\mp}=\frac{\cosh(\xi_{1R}+\phi_1^{1\mp})}{\cosh(\eta_{1R}+\phi_2^{1\mp})}.
\end{eqnarray}  \end{subequations}
Similarly, we derive the intensity expression corresponding to beating soliton $2$, which  is given by
\begin{subequations}
	\begin{eqnarray}
		&&\hspace{-1.5cm}|q_1^{2\mp}|^2=4P_{22}^{\mp}\bigg(k_{2R}^2\cos^2\frac{\theta}{2}P_{12}^{\mp}+l_{2R}^2\sin^2\frac{\theta}{2}(P_{12}^{\mp})^{-1}\label{12a}\\
		&&\hspace{-0.7cm}-k_{2R}l_{2R}\sin\theta\cos[(2\Gamma+k_{2R}^2-l_{2R}^2)z+\theta_1^{2\mp}-\theta_2^{2\mp}-\frac{\pi}{2}]\bigg),\nonumber\\
		&&\hspace{-1.5cm}|q_2^{2\mp}|^2=4P_{22}^{\mp}\bigg(k_{2R}^2\sin^2\frac{\theta}{2}P_{12}^{\mp}+l_{2R}^2\cos^2\frac{\theta}{2}(P_{12}^{\mp})^{-1}\label{12b}\\
		&&\hspace{-0.7cm}+k_{2R}l_{2R}\sin\theta\cos[(2\Gamma+k_{2R}^2-l_{2R}^2)z+\theta_1^{2\mp}-\theta_2^{2\mp}-\frac{\pi}{2}]\bigg),\nonumber
	\end{eqnarray}

where \begin{eqnarray}
&&\hspace{-1.0cm}P_{22}^{\mp}=\frac{1}{D_{22}^{\mp2}}(\cosh(\xi_{2R}+\phi_1^{2\mp})\cosh(\eta_{2R}+\phi_2^{2\mp})),\nonumber\\
&&\hspace{-1.0cm}D_{22}^{\mp}=[\hat{c}_{12}\cosh(\eta_{2R}+\xi_{2R}+\phi_1^{2\mp}+\phi_2^{2\mp}+b_2)\nonumber\\
&&\hspace{0.25cm}+\hat{c}_{12}^{-1}\cosh(\eta_{2R}-\xi_{2R}+\phi_2^{2\mp}-\phi_1^{2\mp}+b_2)], \nonumber\\
&&\hspace{-1.0cm}\text{and}~~ P_{12}^{\mp}=\frac{\cosh(\xi_{2R}+\phi_1^{2\mp})}{\cosh(\eta_{2R}+\phi_2^{2\mp})}.	
\end{eqnarray} \end{subequations}
The expressions (\ref{11a})-(\ref{11b}) and (\ref{12a})-(\ref{12b}) reveal that the linear coupling induces periodic oscillations, arising from the exchange of intensities, which remain unaffected throughout the entire collision process. This is further evidenced by the presence of oscillatory terms both before and after the interaction. Thus, the beating non-degenerate solitons retain their structures during mutual interaction, apart from experiencing a finite phase shift. Another observation from this analysis is that, as in the one-soliton case, the multi-soliton case also shows that the total intensity profiles of the two colliding double-hump solitons exhibit no oscillations and display an elastic collision. To substantiate this, we compute the total intensity without oscillatory term from the asymptotic expressions, which are expressed as
\begin{eqnarray}
	|q_1^{1\mp}|^2+|q_2^{1\mp}|^2=4P_{11}^{\mp}[k_{1R}^2P_{21}^{\mp}+l_{1R}^2(P_{21}^{\mp})^{-1}],\nonumber\\	|q_1^{2\mp}|^2+|q_2^{2\mp}|^2=4P_{22}^{\mp}[k_{2R}^2P_{12}^{\mp}+l_{2R}^2(P_{12}^{\mp})^{-1}]. \label{13}
\end{eqnarray}

We now analyze the interaction dynamics of beating non-degenerate soliton in the presence of oppositely moving degenerate beating soliton to examine their mutual influence on the beating effects. Such a typical collision scenario is illustrated in Figs. \ref{fig5}(b1)-(b3), where the beating effect of the non-degenerate soliton is suppressed, and its original asymmetric double-hump profile re-emerges after the collision. On the other hand, the beating nature of the degenerate soliton is preserved throughout the collision process in both the modes, as a result of the periodic intensity switching accompanied by either enhancement or suppression of its intensity through intensity redistribution. For instance, in Figs. \ref{fig5}(b1)-(b3), the intensity of degenerate beating soliton is enhanced in the first mode $q_1$ whereas it is suppressed in the other mode $q_2$. To facilitate the presence of beating effect of degenerate soliton and the mechanism of intensity redistribution, we have obtained the intensities associated with the asymptotic forms of beating degenerate soliton as
\bes\begin{eqnarray}
&&\hspace{-1.2cm}|q_{j}^{\mp}|^2=\bigg(\cos^2\frac{\theta}{2}|A_{j}^{\mp}|^2+\sin^2\frac{\theta}{2}|A_{k}^{\mp}|^2+(-1)^j\sin\theta|A_j^{\mp}||A_k^{\mp}|\nonumber\\
&&\hspace{0.2cm} \times~ \cos[2\Gamma z+\hat{Q}^{\mp}]\bigg)k_{1R}^2\sech^2(\eta_{1R}+\varphi^{\mp}),\nonumber\\
&&\hspace{0.7cm}~j,k=1,2~(j\neq k),
\end{eqnarray} 
where 	
	\begin{eqnarray}
&&\hspace{-1.3cm}A_{1,2}^-=\frac{\alpha_1^{(1,2)}}{[|\alpha_1^{(1)}|^2+|\alpha_1^{(2)}|^2]^{1/2}},~ A_1^+=\frac{\alpha_1^{(1)}}{\sqrt{|\alpha_1^{(1)}|^2+\chi |\alpha_1^{(2)}|^2}},\nonumber\\ &&\hspace{-1.3cm}A_2^+=\frac{\alpha_1^{(1)}}{\sqrt{|\alpha_1^{(1)}|^2\chi^{-1}+ |\alpha_1^{(2)}|^2}},~ \chi=\frac{|k_1-l_2|^2|k_1+k_2^*|^2}{|k_1-k_2|^2|k_1+l_2^*|^2},\nonumber\\ &&\hspace{-1.3cm}e^{i\hat{Q}^{\mp}}=\frac{A_1^{\mp}A_2^{\mp*}}{A_1^{\mp*}A_2^{\mp}},~ \varphi^-=\frac{1}{2}\ln\frac{|\alpha_1^{(1)}|^2+|\alpha_1^{(2)}|^2}{(k_1+k_1^*)^2},\nonumber\\
&&\hspace{-1.3cm}\text{ and}~~\varphi^+=\frac{1}{2}\ln\frac{|k_1-k_2|^4|k_1-l_2|^2(|\alpha_1^{(1)}|^2+\chi |\alpha_1^{(2)}|^2)}{(k_1+k_1^*)^2|k_1+k_2^*|^4|k_1+l_2^*|^2}.
	\end{eqnarray}\ees
We note that the suppression of intensity switching or periodic oscillations reported in earlier studies on twisted birefringent fibers \cite{radhakrishnan,potasek}, where intensity switching was controlled by appropriately fixing the corresponding complex polarization constants $\alpha_1^{(j)}$, $j = 1, 2$, is entirely distinct from the one occurring above. Hence, the controlling of the spatial oscillation of the beating non-degenerate soliton can be achieved by the intensity redistribution nature of the degenerate beating soliton. To the best of our knowledge, this novel collision property of the beating non-degenerate soliton has not been reported before in the vector solitons literature. Note that this work can be extended to the contexts of nonlinear optics with variable media and BECs with tunable, time-dependent parameters. For example, in nonlinear optical Kerr media with inhomogeneities, the present study can be extended by allowing the nonlinearity coefficient and the linear cross-coupling term $\nu$ in Eq. (\ref{eq1}) to vary along the propagation direction. In the case of BECs, such an extension can be realized by considering tunable linear Rabi coupling and controllable inter- and intra-species interaction strengths \cite{rabi-paper}. The details will be published separately. 

\section{Conclusions}\label{sec-conclu} 
In this paper, we proposed an alternative approach to generate a new class of beating vector solitons. This method relied on the connection between the equations of motion (\ref{eq1}) for two optical modes and the Manakov equation (\ref{manakov}). Then, by superimposing the non-degenerate vector soliton solutions of the latter integrable CNLS equations and in the presence of intensity switching between the modes we were able to achieve the desired beating non-degenerate vector solitons. We found that these solitons exhibit elastic collision with sustained beating effects when interacting with themselves. Their beating nature can be controlled by allowing them to collide with degenerate beating solitons. We have analysed the underlying controlling mechanism through suitable asymptotic analysis. The results presented in this work will provide new insights into beating solitons in BECs, nonlinear optics, and related areas of research.

\noindent{\bf Acknowledgement}\\
M.L. thanks DST-ANRF, INDIA for the award of a DST-ANRF National Science Chair  (NSC/2020/000029) position in which S.S. is a Research Associate.


\noindent{\bf Declaration of Competing Interest}\\ The authors declare that they have no known competing financial interests or personal relationships that could have appeared to influence the work reported in this paper.


\setstretch{01.20}

\end{document}